# A METHOD FOR EXTRACTING MAXIMUM RESOLUTION POWER SPECTRA FROM GALAXY SURVEYS[1]

Max Tegmark

*Max-Planck-Institut für Physik, Föhringer Ring 6, D-80805 München*
*Max-Planck-Institut für Astrophysik, Karl-Schwarzschild-Str. 1, D-85740 Garching*
*max@mppmu.mpg.de*

## Abstract

The power spectrum estimated from a galaxy redshift survey is the real spectrum convolved with a window function, so when estimating the power on very large scales (for small $k$), it is important that this window function be as narrow as possible. A method that achieves this is presented. The optimal fluctuation estimate is found to be the Fourier transform of the number density fluctuations $n/\bar{n} - 1$ weighted by a function $\psi_0$, and the optimal $\psi_0$ is found to be the ground-state solution of the Schrödinger equation, with the inverse selection function as the potential. This quantum mechanics analogy occurs basically because we want the weight function to be narrow both in Fourier space (to give a narrow window) and in real space (to minimize the variance from shot noise). An optimal method for averaging the estimates at different **k** is also presented, generalizing the standard procedure of averaging over shells in **k**-space. Finally, a discrete version of the method is presented, dividing space into "fuzzy pixels", which has the advantage of being able to handle redshift distortions in a straightforward way.

---

# 1 Introduction

Our observational knowledge of the large-scale structure of the universe is growing at a formidable rate. Fifteen years ago, redshifts had been measured for a few thousand galaxies. Today the figure is around $10^5$, and ongoing projects such as the AAT 2dF Survey and the Sloan Digital Sky Survey will raise it to $10^6$ within a few years. It is thus very timely to develop analysis techniques that allow us to make the most of this data. One of the most important cosmological quantities that we wish to extract from redshift surveys is the power spectrum $P(k)$. Although this function has been estimated from a wide variety of galaxy samples, selected in optical (Baumgart & Fry 1991; Park *et al.* 1992; Vogeley *et al.* 1992; Park *et al.* 1994, Vogeley 1994 – hereafter V94), infra-red (Fisher *et al.* 1993 – hereafter F93; Feldman, Kaiser & Peacock 1994 – hereafter FKP) and radio (Peacock & Nicholson 1991) frequencies, there is still some disagreement about both its overall normalization and its behavior on very large scales. The former problem is related to the issue of biased galaxy formation – see Peacock & Dodds (1994) and references therein. Sorting out the behavior on very large scales, however, is more an issue of survey geometry and data analysis. FKP find evidence for $P(k)$ turning over, whereas Park *et al.* (1992) claim a continued rise up to $200h^{-1}$Mpc. As pointed out by many authors (*e.g.* F93, FKP, V94, Efstathiou 1994), the $P(k)$ we estimate is the true power spectrum convolved with some window function, and the differences between the window functions of various workers may be part of the explanation for the discrepancies.

The window functions depend on both the survey geometry and the analysis technique employed. Although we may wish we had a method where the window functions were delta functions, it is easy to see that this is impossible given a finite survey volume. Given this constraint, it is natural to ask which analysis technique gives the *narrowest* window functions. The purpose of the present paper is to answer this question.

Before embarking on this program, it is instructive to compare this approach with that of two other recently published methods. Both the method of FKP and the Karhunen-Loève eigenmode method of V94 are optimal in the sense of maximizing the signal-to-noise ratio within a certain class of techniques, without specifically paying attention to the width of the window functions[2]. This corresponds to minimizing the *vertical* error bars that

---
[2]The latter method nonetheless tends tend to sharpen the window functions, although

are conventionally placed on power spectrum estimates. To indicate the width of the window functions, *i.e.*, the extent of spectral blurring, one can also place *horizontal* error bars on the estimates. This has recently become fashionable when plotting estimates of the cosmic microwave background power spectrum from various experiments, and would be welcome also in the context of galaxy surveys. The key point to remember is that if we are concerned about both vertical and horizontal error bars, then *there is no one optimal method*. The reason is that there is a trade-off between the two. As pointed out in *e.g.* F93, if one places more weight on the most distant galaxies in a flux-limited survey, then the horizontal error bars get narrower but the vertical ones grow, because of increased shot noise contributions. A method is clearly superior to another if it produces smaller error bars *both* horizontally and vertically. We will find a method that is unbeatable in this sense, with a free parameter $\gamma$ that allows the desired ratio of the two error bars to be specified by the user.

It should be emphasized that the concept of window functions and "horizontal error bars" (under various names) and the advantage of them being narrow has been extensively discussed in the prior literature (*e.g.* FKP, F93, V94). V94 even provides explicit plots of the window functions for some specific examples. Thus what is new in the present paper is mainly the derivation of a method that explicitly optimizes the window functions.

This paper is organized as follows. In Section 2, we derive the rank one estimate of $P(\mathbf{k})$ that is optimal in the above sense. In Section 3, we give examples for a number of common survey configurations, and attempt to provide some physical intuition for how the geometry affects our ability to estimate $P(k)$. In Section 4, we generalize the traditional averaging over shells in $\mathbf{k}$-space by deriving the optimal way to combine the estimates for different $\mathbf{k}$-modes. In Section 5, we present a pixelized version of the treatment, which has the advantage of readily being able to handle additional elements of realism such as redshift-space distortions. The conclusions are summarized in Section 6, and some mathematical details are given in the Appendix.

---

not optimally, as a kind of side effect, indicating that narrow windows are "natural" in some information-theoretic sense.

## 2 The optimal weight function

Just as FKP, we model the observed galaxy distribution as a 3D Poisson process $n(\mathbf{r}) = \sum_i \delta(\mathbf{r}-\mathbf{r}_i)$ with intensity $\lambda(\mathbf{r}) = \bar{n}(\mathbf{r})[1+\delta_r(\mathbf{r})]$. The function $\bar{n}$ is the selection function of the galaxy survey under consideration. As is customary, we model the density fluctuations $\delta_r$ as a Gaussian random field with power spectrum $P(\mathbf{k}) = P(k)$. We estimate the fluctuation amplitude at wave vector $\mathbf{k}^*$ by[3]

$$\widehat{F}(\mathbf{k}^*) \equiv \int \varphi(\mathbf{r}) n(\mathbf{r}) d^3r \tag{1}$$

and wish to find the optimal choice of the weight function $\varphi$. (In Section 4, we will see that this *Ansatz* is essentially unbeatable, since all quadratic power estimators lead back to this one.) In the Appendix, we show that

$$\langle \widehat{F}(\mathbf{k}^*) \rangle = \widehat{\psi}(0), \tag{2}$$

$$\langle |\widehat{F}(\mathbf{k}^*)|^2 \rangle = |\widehat{\psi}(\mathbf{0})|^2 + \int |\widehat{\psi}(\mathbf{k})|^2 P(\mathbf{k}) d^3k + \int |\psi(\mathbf{r})|^2 \frac{1}{\bar{n}(\mathbf{r})} d^3r. \tag{3}$$

This corresponds to equation (2.1.6) of FKP. Hats denote Fourier transforms, and we have defined $\psi(\mathbf{r}) \equiv \bar{n}(\mathbf{r})\varphi(\mathbf{r})$. From here on, we will find it convenient to use the standard Dirac quantum mechanics notation with kets and bras, where as usual the wave number operator $\widehat{\mathbf{k}} = -i\nabla$. Defining our raw estimate of the power at wave vector $\mathbf{k}$ as

$$Q(\mathbf{k}) \equiv |\widehat{F}(\mathbf{k})|^2 - |\widehat{\psi}(\mathbf{0})|^2, \tag{4}$$

we can write equation (3) simply as $\langle Q(\mathbf{k}) \rangle = \langle \psi | P(\widehat{\mathbf{k}}) + V(\widehat{\mathbf{r}}) | \psi \rangle$, where we have defined the "potential" function $V(\mathbf{r}) \equiv 1/\bar{n}(\mathbf{r})$. Thus our power estimate is a sum of two terms: a weighted average of the cosmological power $P(\mathbf{k})$, the weights being the window function $|\widehat{\psi}(\mathbf{k})|^2$, and a contribution from shot noise. Since the integral of the window function should be unity, the correct normalization of $\psi$ is evidently $\langle \psi | \psi \rangle = 1$. Subtracting the shot noise contribution, we obtain the unbiased power estimate $Q(\mathbf{k}) - \langle \psi | V(\mathbf{r}) | \psi \rangle$. In the approximation of FKP that $\widehat{F}(\mathbf{k})$ is Gaussian,

---

[3]Including a random mock survey as in equation (2.1.3) in FKP can never give minimal error bars, since inclusion of random numbers will always increase the variance of the estimator.

the variance of this estimate is simply[4] $\langle|\widehat{F}(\mathbf{k})|^4\rangle - \langle|\widehat{F}(\mathbf{k})|^2\rangle^2 = \langle Q(\mathbf{k})\rangle^2$. In other words, we minimize the vertical error bars on our power estimate by minimizing $\langle Q(\mathbf{k})\rangle$. Since for any reasonably narrow window function, the cosmic contribution $\langle\psi|P(\widehat{\mathbf{k}})|\psi\rangle \approx P(\mathbf{k}^*)$, independent of $\psi$, this means that we simply want to minimize the shot noise contribution $\langle\psi|V(\widehat{\mathbf{r}})|\psi\rangle$.

Let us summarize our results so far: a good $\psi$ should be narrow both in Fourier space (to give a narrow window function, *i.e.*, small horizontal error bars) and in real space (to minimize the impact of shot noise and thus give small vertical error bars). In other words, we arrive at the following optimization problem: minimize $\langle\psi|p(\widehat{\mathbf{k}}) + V(\widehat{\mathbf{r}})|\psi\rangle$ subject to the constraint that $\langle\psi|\psi\rangle = 1$. Here $p$ is some *penalty function* that we choose to be small near the $\mathbf{k}^*$ which we wish to estimate and large elsewhere. Introducing a Lagrange multiplier $E$, this leads to the eigenvalue equation

$$[p(\widehat{\mathbf{k}}) + V(\widehat{\mathbf{r}})]|\psi\rangle = E|\psi\rangle. \tag{5}$$

Throughout this paper, we will estimate the power at wave vector $\mathbf{k}^*$ by chosing the quadratic penalty function $p(\mathbf{k}) = |\mathbf{k} - \mathbf{k}^*|^2/2\gamma$, *i.e.*, try to make the window function narrow in the least squares sense. By direct substitution, it is readily seen that equation (5) has the solution

$$\psi(\mathbf{r}) = \psi_0(\mathbf{r})e^{-i\mathbf{k}^*\cdot\mathbf{r}}, \tag{6}$$

where $\psi_0$ is the solution to the Schrödinger equation

$$\left[-\frac{1}{2}\nabla^2 + \gamma V(\mathbf{r})\right]\psi_0 = E\psi_0 \tag{7}$$

with the smallest eigenvalue $E$, the ground state. Continuing our quantum analogy, we see that we arrived at the quantum ground state because we wanted the normalized $\psi_0$ that minimized the total "energy", where the "kinetic energy" $\langle\psi_0|\widehat{\mathbf{k}}^2|\psi_0\rangle/2$ corresponded to the horizontal error bar and the "potential energy" $\gamma\langle\psi_0|V(\widehat{\mathbf{r}})|\psi_0\rangle$ corresponded to the vertical error bar. In conclusion, we see that we can rewrite equation (1) as

$$\widehat{F}(\mathbf{k}^*) \equiv \int e^{-i\mathbf{k}^*\cdot\mathbf{r}}\psi_0(\mathbf{r})\frac{n(\mathbf{r})}{\bar{n}(\mathbf{r})}d^3r. \tag{8}$$

---

[4] This follows from the fact that the real and imaginary parts have the same variance, which in turn follows from the assumption of random phases. A real-valued $\widehat{F}(\mathbf{k})$ would gives twice this value. Since a translation in real space corresponds to a phase modulation in Fourier space, the random phases assumption follows directly from assuming that the statistical properties of the galaxy distribution are translationally invariant, an assumption that of course underlies the entire power spectrum formalism.

In other words, although there was no mention of Fourier transforms is our *Ansatz* (1), our optimal power estimator $\widehat{F}(\mathbf{k}^*)$ is simply the Fourier transform of the field $F(\mathbf{r}) = \psi_0(\mathbf{r})\frac{n(\mathbf{r})}{\bar{n}(\mathbf{r})}$. Since $n$ is just a sum of delta functions, this reduces to

$$\widehat{F}(\mathbf{k}^*) = \sum e^{-i\mathbf{k}^*\cdot\mathbf{r}_i}\frac{\psi_0(\mathbf{r}_i)}{\bar{n}(\mathbf{r}_i)}, \tag{9}$$

where the sum is to be taken over all galaxies observed in the survey.

## 3  Examples

In this section, we give the optimal weight function $\psi_0$ for a few selected survey geometries, and discuss some of their properties. A general feature of all solutions is seen to be that they fall off smoothly near the survey boundaries, and tend to assign very low weights to points near corners and sharp edges. Basically, this is because it helps a lot in Fourier space without costing much in real space.

### 3.1  Volume-limited surveys

For a volume-limited survey, $\bar{n}(\mathbf{r})$ equals some constant when $r$ is in the survey volume, and vanishes otherwise. This corresponds to the potential $V(\mathbf{r})$ being infinite outside the survey volume, so we can without loss of generality set $V(\mathbf{r}) = 0$ inside[5] and obtain the quantum-mechanical particle-in-the-box problem: $\psi_0$ must satisfy the Helmholtz equation $[\nabla^2 + 2E]\psi = 0$ inside the region and vanish on its boundary. For a volume-limited all sky survey complete out to a radius $R$, we obtain

$$\psi_0(\mathbf{r}) \propto j_0\left(\frac{\pi r}{R}\right), \tag{10}$$

where $j_0(x) = (\sin x)/x$. If we approximate the volume in a pencil-beam survey by a cylinder as in Kaiser & Peacock (1991), with radius $R$, length $L$ and cylindrical coordinates $\rho \equiv [x^2 + y^2]^{1/2} < R$, $0 < z < L$, we obtain

$$\psi_0(\mathbf{r}) \propto \sin(\pi z/L) J_0(k_0\rho/R), \tag{11}$$

---

[5] Continuing our quantum analogy, we get $V(\mathbf{r}) = 0$ by subtracting the constant energy $\gamma/\bar{n}$ from the Hamiltonian in brackets in equation (7). As we know, such a change in the (arbitrary) zero point makes no physical difference: all eigenvalues $E$ will simply decrease by $\gamma/\bar{n}$, and the eigenfunctions will remain the same.

where $k_0 \approx 2.405$ and $J_0$ is the zeroth Bessel function. A more realistic pencil beam survey contains a *conical* volume of opening angle $2\theta_1$, in spherical coordinates defined by $\theta < \theta_1$, $r < R$. Both this case and that of a *slice* $\theta_0 < \theta < \theta_1$, $\varphi_0 < \varphi < \varphi_1$, $r < R$ are treated in Pockels (1891). The solutions are a spherical Bessel function of non-integer order in the radial direction, an a non-integer spherical harmonic for the angular part. We will return to these solutions in subsection 3.3 below.

## 3.2 Flux-limited and diameter-limited surveys

For a flux-limited survey, the selection function $\bar{n}(\mathbf{r})$ has the radial dependence (*e.g.* Peebles 1993, p214)

$$\bar{n}(r) \propto \int_{4\pi f r^2}^{\infty} \phi(L) dL, \tag{12}$$

where $\phi$ is the galaxy luminosity function and $f$ the flux limit of the survey. The Schechter luminosity function $\phi(L) \propto L^\alpha e^{-L/L_*}$, $\alpha = -1.07 \pm 0.05$, gives $\bar{n}(\mathbf{r}) = \Gamma(1 + \alpha, 4\pi f r^2/L_*)$, an incomplete Gamma function. In the crude approximation $\phi(L) \approx L^{-2}$, we obtain $\bar{n}(\mathbf{r}) \propto r^{-2}$. Thus $V(\mathbf{r}) \propto r^2$ for an all-sky survey, and the Schrödinger equation (7) becomes that of the harmonic oscillator, whose familiar ground state is

$$\psi_0(\mathbf{r}) \propto e^{-(r/R)^2/2}, \tag{13}$$

a Gaussian. Here $R \propto \gamma^{-1/4}$, where $\gamma$ determined the height of the potential in equation (7). Thus we see that the parameter $\gamma$, which specifies how concerned we are about shot noise relative to our desire to keep the window function narrow, determines the effective depth of the survey. A greater depth of course narrows the window function, at the price of more shot noise. For the Schechter model, $V(\mathbf{r})$ grows exponentially for large $r$, which causes the resulting $\psi_0$ to fall off even faster than a Gaussian at large radii[6]. Here the effect of changing $\gamma$ is found to be much smaller, as this exponential cutoff defines a survey depth rather naturally. Notice that in the extreme case of volume-limited surveys discussed above, where the survey depth was completely fixed, the dependence of $\psi_0$ on $\gamma$ vanished entirely.

For diameter-limited surveys, the calculation of $\bar{n}$ is completely analogous, with the angular distribution function replacing the luminosity function in equation (12).

---

[6]It is easy to see that $|\psi_0|^2$ will indeed always fall off faster than $1/r^3 V(r)$, for otherwise the "potential energy" $\langle \psi | V(\widehat{\mathbf{r}}) | \psi \rangle$, which is the expected shot noise contribution, would be infinite.

## 3.3 Combinations: realistic pencil beams, slices, etc.

Most realistic surveys are volume-limited in the angular directions (due to incomplete sky coverage) and flux-limited or diameter-limited in the radial direction. They thus result in a Schrödinger equation with a purely radial potential, supplemented with the boundary conditions that $\psi$ must vanish on the survey boundaries, as discussed in subsection 3.1 above. This problem is readily attacked by separation of variables in spherical coordinates, writing the weight function as

$$\psi_0(\mathbf{r}) = R(r)\Theta(\theta)\Phi(\varphi). \tag{14}$$

Let us restrict our attention to the following two common cases: a *pencil beam* where $\bar{n} = 0$ outside the cone $\theta < \theta_1$ and a $\Delta\theta \times \Delta\varphi$ *slice* where $\bar{n} = 0$ outside the region $|\varphi| < \Delta\varphi/2$, $|\theta - \pi/2| < \Delta\theta/2$. (A slice which is not centered on the equator can of course be put in this form by rotating the coordinate system.) The ground state solutions for $\Theta$ and $\Phi$ are the non-integer spherical harmonics given by Pockels (1891). When the survey subtends merely a small angle in the sky ($\ll$ one radian), these angular solutions reduce to

$$\Theta(\theta)\Phi(\varphi) \approx J_0\left(\frac{k_0\theta}{\theta_1}\right) \tag{15}$$

for the pencil beam case and

$$\Theta(\theta)\Phi(\varphi) \approx \cos\left(\frac{\pi\varphi}{\Delta\varphi}\right)\cos\left[\frac{\pi}{\Delta\theta}\left(\theta - \frac{\pi}{2}\right)\right] \approx \cos\left(\frac{\pi y}{\Delta\varphi}\right)\cos\left(\frac{\pi z}{\Delta\theta}\right) \tag{16}$$

for the case of the slice, which is readily shown by direct substitution into the standard expression for $\nabla^2$ in spherical coordinates. The corresponding angular eigenvalues, which we denote $E_\Omega$, are $E_\Omega = (k_0/\theta_1)^2$ for the pencil beam and $E_\Omega = (\pi/\Delta\theta)^2 + (\pi/\Delta\varphi)^2$ for the slice. ($k_0 \approx 2.405$ as before.) If the exact results of Pockels are used for the angular part, these eigenvalues will of course be slightly different.

Inserting this into equation (7), we obtain the following ordinary differential equation for the radial function $R(r)$:

$$(rR)'' = \left[\frac{E_\Omega}{r^2} + 2\gamma V(r) - 2E\right]rR. \tag{17}$$

Assuming that $V(r)$ remains finite at $r = 0$, the first term on the right hand side will dominate near the origin, giving a power-law solution

$$R(r) \propto r^\alpha \tag{18}$$

for small $r$. Substituting this into equation (17) and requiring $\psi$ to remain finite at the origin ($\alpha \geq 0$), we obtain

$$\alpha = \frac{\sqrt{1 + 4E_\Omega} - 1}{2}. \qquad (19)$$

Equation (17) is now readily solved numerically by the "shooting" method, using equation (18) as initial data and integrating the differential equation for various choices of the constant $E$. The correct ground-state eigenvalue $E$ is of course the one for which the resulting solution $R(r)$ approaches zero as $r \to \infty$ and has no zero-crossings — see *e.g.* Hajj *et al.* 1974 for details.

If a good approximation is considered satisfactory, the standard variational method (making an *Ansatz* with a few free parameters and minimizing $\langle\psi|H|\psi\rangle/\langle\psi|\psi\rangle$) is a recommended alternative approach, as all one needs is the ground state.

Regardless of the exact form of the radial selection function, we can draw a general conclusion from equations (18) and (19): $\psi$ shuns sharp corners. For instance, for a slice with $\Delta\theta = 20°$ and $\Delta\varphi = 30°$, we get $\alpha \approx 10$, so that $\psi \propto r^{10}$ for small $r$, assigning almost no weight at all to the galaxies closest to us, in the sharp corner of the slice. A pencil beam with an opening angle of $2°$, *i.e.*, with $\theta_1 = 1°$, is even more extreme, giving $\alpha \approx 137$.

Note that for several identical, non-intersecting volumes (such as pencil beams), the ground state is degenerate (just as in the corresponding quantum problem). Thus there will be one solution $\psi_1$ which is nonzero only in the first "pencil", another equally good solution $\psi_2$ which is nonzero only in the second "pencil" and is just a translated and rotated version of $\psi_1$, *etc*. These solutions give $E$ just as low as a combination like $(\psi_1 + \psi_2)/\sqrt{2}$. This leads to the perhaps surprising conclusion that two pencil beams cannot give narrower window functions than one[7]. Notwithstanding, it is of course worthwhile to make several independent pencil beam surveys, since averaging different modes as in Section 4 helps reduce both the variance from shot noise and the sample variance.

### 3.4 The k-space picture

Figure 1 is an attempt to clarify what is going on in **k**-space. The funnel-shaped surface is the Cold Dark Matter (CDM) power spectrum $P(\mathbf{k}) =$ $P(\sqrt{k_x^2 + k_y^2 + k_z^2})$ of Bond and Efstathiou (1984), with $h\Omega = 0.5$ and $k$ measured in units of $(h^{-1}\text{Mpc})^{-1}$. If the radial profile looks unfamiliar, it is because the axes are linear rather than logarithmic. For clarity, the third coordinate has been suppressed in the plot ($k_z = 0$). The two surfaces below are examples of window functions. The estimated power is obtained by simply multiplying the window function with the "funnel" and integrating over all **k**. We have seen that by changing $\mathbf{k}^*$ in equation (8), we can slide our window function around and center it at a point of our choice in **k**-space, but that its shape will remain unchanged, being simply $|\widehat{\psi}_0|^2$. This simply reflects the well-known fact that the spectral resolution $\Delta k$ is independent of **k**. Thus the principal difficulty is to estimate the power at very long wavelengths, near the origin $\mathbf{k} = 0$, where $P(\mathbf{k})$ may change dramatically on the scale of the width of the window function. To accurately probe the largest scales $\lambda = 2\pi/k \gg 30\text{Mpc}$, where $P(k) \propto k$, we thus need a window function that is so narrow that it "fits inside the funnel". Although narrower window functions can always be achieved with deeper surveys (wider $\psi_0$), the two window functions plotted qualitatively illustrate two subtle problems that can occur even if the survey depth $L \gg 1/k$. The width (standard deviation) of $\psi_0$ determines the scale on which the central peak of $\widehat{\psi}_0$ falls off. Since the example on the left has a very narrow central peak, we see that it indeed corresponds to a very deep survey. Unfortunately, the "ringing" in Fourier space (often referred to as sidelobes) spoils things, since even if we place the central peak in the middle of the funnel, the dominant contribution will come from the sidelobes picking up power from much larger $k$. Such sidelobes are caused by sharp edges in $\psi_0$, caused for instance by incomplete sky coverage, and it is precisely this problem that the method presented in this paper attempts to minimize.

---

[7] This result is not at odds with the "cross-correlation" argument described in V94. The latter combines several different modes, as is done in the following section, but in such a way that the power estimate is not positive definite. Thus the "cross-correlation" window function $W(k)$ can take negative values, which renders its usefulness unclear.

The second window function example, on the right, has a different problem. Although $\psi_0$ is evidently smooth (as there are no sidelobes), the geometry is highly anisotropic. This is qualitatively what happens for a pencil beam (where $|\widehat{\psi}_0|^2$ will be shaped like a thin disc in **k**-space) and a thin "slice" (where $|\widehat{\psi}_0|^2$ will be a pencil in **k**-space). Thus even though the window function $|\widehat{\psi}_0|^2$ is narrow enough in the most favorable direction (because the survey is deep enough in one or two dimensions), it will still not "fit inside the funnel". Rather, the power estimate will be dominated by the ends of the window function protruding out into parts of **k**-space where $k$ and $P(k)$ are much larger. This well-known phenomenon has been emphasized by *e.g.* Kaiser & Peacock (1991), who argued that the apparent power excess at $128h^{-1}$ Mpc in a pencil beam survey reported by Broadhurst *et al.* 1990 was probably due to such leakage from shorter wavelengths. In summary, the ability for a survey to probe the longest wavelengths is limited by its weakest link, the dimension in which it is the narrowest.

Figure 2 gives a quantitative example of sidelobe reduction for the above-mentioned case of a volume-limited, all-sky survey of depth $R = 300h^{-1}$Mpc. Since the main difficulty is very small $k$, we plot the most large-scale window function attainable, that devised to estimate $P(0)$. The optimal $\psi_0$ from equation (10) gives the window function

$$W(k) = 4\pi \left(\frac{\sin x}{\pi^2 - x^2}\right)^2, \qquad (20)$$

where $x \equiv kR$, whereas the naive choice $\psi_0$ constant gives

$$W(k) = \frac{6}{\pi x^2}\left(\frac{\sin x}{x} - \cos x\right)^2. \qquad (21)$$

They have both been correctly normalized, to integrate to unity. They both vanish at $k = 0$ for phase space reasons – $P(0)$ is of course theoretically unobservable. For comparison, two unnormalized power spectra, whose shapes one would like to be able to discriminate between, are plotted; the CDM spectrum of Figure 1 and the a Baryonic Dark Matter (BDM) spectrum. The latter (Hu 1995) has $\Omega = 0.2$, $h = 0.8$ and a thermal history where the universe remained 10% ionized.

## 4 The optimal weighting of Fourier modes

So far, we have discussed how to estimate the power $P(\mathbf{k})$ at any given wave vector $\mathbf{k}$. The galaxy clustering is generally believed to be isotropic, which means that $P$ in fact depends only on $k = |\mathbf{k}|$, the magnitude of the wave vector. The conventional way to estimate $P(k)$ is to average the estimates of $P(\mathbf{k})$ over a fairly thin spherical shell in $\mathbf{k}$-space of radius $k$, thereby reducing the vertical error bar. We will now examine more general weighting schemes.

### 4.1 The most general quadratic estimator

It is easy to see that the most general power estimator whose expectation value is linear in $P(\mathbf{k})$ must be quadratic in the field quantities, *i.e.*, of the

form
$$Q(\mathbf{k}^*) = \int A(\mathbf{r},\mathbf{r}')\frac{n(\mathbf{r})n(\mathbf{r}')}{\bar{n}(\mathbf{r})\bar{n}(\mathbf{r}')}d^3r d^3r' - B(\mathbf{k}^*) \tag{22}$$

— estimators with higher-order terms would give quantities like $P(\mathbf{k})^2$ in the expectation values, and be fairly useless assuming that $\delta_r$ is Gaussian. The functions $A$ and $B$ are arbitrary. We lost no generality by inserting $\bar{n}$ above, as these functions could always be absorbed into $A$. It is straightforward to show that

$$\langle Q(\mathbf{k}^*)\rangle = \int W(\mathbf{k})P(\mathbf{k})d^3k + b(\mathbf{k}^*) - B(\mathbf{k}^*), \tag{23}$$

*i.e.*, that the expectation value of the estimator is always just the power spectrum convolved with some window function $W$ plus a bias term $b$ from shot noise *etc.* that is independent of $P(\mathbf{k})$. Thus one should simply choose $B(\mathbf{k}^*) = b(\mathbf{k}^*)$ to get an unbiased estimator. The interesting part is how to choose the function $A$. Since $Q(\mathbf{k}^*)$ must be real-valued, we can without loss of generality choose $A$ to be Hermitean, *i.e.* $A(\mathbf{r},\mathbf{r}') = A(\mathbf{r}',\mathbf{r})^*$. This means that, viewed as an operator, all its eigenvalues $w_i$ are real and it can be expanded in a set of orthogonal eigenfunctions as

$$A(\mathbf{r},\mathbf{r}') = \sum_i w_i \psi_i(\mathbf{r})\psi_i(\mathbf{r}')^*. \tag{24}$$

Thus equation (22) becomes

$$Q(\mathbf{k}^*) = \sum w_i \left|\int \psi_i(\mathbf{r})\frac{n(\mathbf{r})}{\bar{n}(\mathbf{r})}d^3r\right|^2 - B(k), \tag{25}$$

which we recognize as simply a linear combination of estimates of the type studied in Section 2 – see equations (1) and (4). Normalizing the the functions as $\langle\psi_i|\psi_i\rangle = 1$, as is Section 2, $Q(\mathbf{k}^*)$ must indeed be simply a weighted average, where $\sum w_i = 1$. To guarantee that the window function $W$ be non-negative, we will require $w_i > 0$, *i.e.* that $A$ be positive definite. Resuming the quantum analogy, we can thus think of the most general $A$ as a density operator

$$A = \sum_i w_i|\psi_i\rangle\langle\psi_i|, \tag{26}$$

and write the normalization and expectation value equations as simply tr$A = 1$ and $\langle Q(\mathbf{k}^*)\rangle = \text{tr}\{[P(\widehat{\mathbf{k}}) + V(\widehat{\mathbf{r}})]A\}$, respectively.

Thus hopefully having demystified the notion of the most general power estimator slightly, showing that it is really nothing more than a weighted average of simple estimates estimates like in equation (4), it is clear that we want to split our analysis into two steps: first estimate the power at a grid of points in **k**-space, with as narrow window functions as possible, then for each $k$-value of interest, average these estimates in some clever way to reduce the variance. We dealt with the first step in Section 2, so let us now take the function $\psi_0$ as given, and find the optimum way to average the different Fourier modes.

## 4.2 The optimal weights

We define the *covariance function*

$$C(\mathbf{k},\mathbf{k}') \equiv \langle \widehat{F}(\mathbf{k})^*\widehat{F}(\mathbf{k}')\rangle - \langle \widehat{F}(\mathbf{k})\rangle^*\langle \widehat{F}(\mathbf{k})\rangle. \tag{27}$$

Let us split it into a sum of two contributions, one from cosmic power and one from shot noise, as $C(\mathbf{k},\mathbf{k}') = C_c(\mathbf{k},\mathbf{k}') + C_s(\mathbf{k},\mathbf{k}')$. In the Appendix, we show that $\langle \widehat{F}(\mathbf{k})\rangle = \widehat{\psi}_0(\mathbf{k})$ and

$$C_c(\mathbf{k},\mathbf{k}') = \int \widehat{\psi}_0(\mathbf{k}''-\mathbf{k}')^*\widehat{\psi}_0(\mathbf{k}''-\mathbf{k})P(\mathbf{k}'')d^3k'', \tag{28}$$

$$C_s(\mathbf{k},\mathbf{k}') = \int e^{-i(\mathbf{k}'-\mathbf{k})\cdot\mathbf{r}}|\psi_0(\mathbf{r})|^2\frac{1}{\bar{n}(\mathbf{r})}d^3r. \tag{29}$$

This corresponds to equation (2.2.2) in FKP, with their function $S$ giving the shot noise. Since the functions $\psi_0$ and $|\psi_0|^2/\bar{n}$ both have widths of the order of the size of the survey volume $L$, their Fourier transforms tend to fall off on a *coherence scale* $\Delta k \sim 1/L$ — see FKP. Assuming that $P(\mathbf{k})$ does not vary much when **k** changes by such a small amount, we can factor it out of equation (28), apply the convolution theorem and obtain

$$C_c(\mathbf{k},\mathbf{k}') \approx P\left(\frac{\mathbf{k}+\mathbf{k}'}{2}\right)\int e^{-i(\mathbf{k}'-\mathbf{k})\cdot\mathbf{r}}|\psi_0(\mathbf{r})|^2 d^3r. \tag{30}$$

Note that for volume-limited surveys, where $\bar{n}(\mathbf{r})$ is constant, $C_c$ and $C_s$ differ merely by a factor $\bar{n}P(k)$. Also note that for a highly anisotropic survey volume, such as a pencil-beam or a slice, the coherence length is correspondingly anisotropic in **k**-space, being shortest in the direction corresponding to the greatest dimension of the survey volume. In other words, the coherence behaves in much the same way as the window function $|\widehat{\psi}_0(\mathbf{k})|^2$, as in the lower right of Figure 1.

Suppose we have evaluated the $Q(\mathbf{k})$ of equation (4) at the $N$ different $\mathbf{k}$-values $\{\mathbf{k}_1, ..., \mathbf{k}_N\}$, and arranged the shot-noise corrected results in an $N$-dimensional vector $\mathbf{q}$, *i.e.*,

$$q_i \equiv Q(\mathbf{k}_i) - \int |\psi_0(\mathbf{r})|^2 \frac{1}{\bar{n}(\mathbf{r})} d^3r. \tag{31}$$

These points $\mathbf{k}_i$ are preferably chosen in the vicinity of the sphere $|\mathbf{k}| = k^*$, and it is clearly redundant to use points separated by much less than the coherence length $\Delta k$, as this adds almost no additional information.
Let us define another $N$-dimensional vector, $\mathbf{e}$, that has all its components equal to one, *i.e.*, $e_i \equiv 1$. As our estimate $\tilde{P}(k)$ of the true power $P(k)$, we take a weighted average $\tilde{P}(k) \equiv \mathbf{w} \cdot \mathbf{q}$, where the weights $w_i \geq 0$ add up to one, *i.e.*, $\mathbf{e} \cdot \mathbf{w} = 1$. The expectation value of our estimate is simply

$$\langle \tilde{P}(k) \rangle = \int W(\mathbf{k}) P(k) d^3k, \tag{32}$$

where

$$W(\mathbf{k}) \equiv \sum_{i=1}^{N} w_i |\widehat{\psi}(\mathbf{k} - \mathbf{k}_i)|^2, \tag{33}$$

and its variance is

$$\langle \tilde{P}(k)^2 \rangle - P(k)^2 = \frac{1}{2} \mathbf{w}^t C \mathbf{w}, \tag{34}$$

where the matrix $C$ is given by $C_{ij} = 4|C(\mathbf{k}_i, \mathbf{k}_j)|^2$. Just as in FKP, we are of course forced to make some assumption about the power level $P(k)$ to be able to compute the function $C$ and estimate our error bars.

The variance clearly decreases if we make the distribution $w_i$ less concentrated, effectively averaging more independent modes. For an all-sky survey with no zones of avoidance, $\psi_0$ and $\widehat{\psi}_0$ will be spherically symmetric, so averaging over a shell in $\mathbf{k}$-space will not widen the window function $W(\mathbf{k})$ at all. However, this is not a very common survey geometry, and we may in addition wish to do some averaging in the radial direction to further reduce the variance, while still striving to keep the window function fairly narrow. As in the last section, we attempt to minimize a penalty

$$\int p(k) W(\mathbf{k}) d^3k = \mathbf{f} \cdot \mathbf{w}, \tag{35}$$

where $f_i \equiv f(\mathbf{k}_i)$ and the function $f$ works out to be the convolution of $|\widehat{\psi}|^2$ with $p$:

$$f(\mathbf{k}) = \int |\widehat{\psi}(\mathbf{k}' - \mathbf{k})|^2 p(k') d^3k'. \tag{36}$$

Note that this time the penalty function $p$ depends only on the magnitude $k = |\mathbf{k}|$.

We thus arrive at the following optimization problem:
Minimize $\mathbf{w}^t C \mathbf{w}/2 + \mathbf{f} \cdot \mathbf{w}$ subject to the constraints that $\mathbf{e} \cdot \mathbf{w} = 1$ and $w_i \geq 0$. This is a quadratic programming problem. However, there is hardly much point in delving into complicated numerical methods at this step, as we merely want a decent solution and do not care if it is *exactly* optimal. In this spirit, it is quite easy to obtain an approximate solution as follows. Introducing a Lagrange multiplier $\lambda$, we minimize $\mathbf{w}^t C \mathbf{w}/2 + \mathbf{f} \cdot \mathbf{w} - \lambda \mathbf{e} \cdot \mathbf{w}$ and obtain $C\mathbf{w} = \lambda \mathbf{e} - \mathbf{f}$. Eliminating $\lambda$ using $\mathbf{e} \cdot \mathbf{w} = 1$, we get

$$\mathbf{w} = \left(\frac{1 + \mathbf{e} \cdot \mathbf{f}'}{\mathbf{e} \cdot \mathbf{e}'}\right) \mathbf{e}' - \mathbf{f}', \tag{37}$$

where $\mathbf{e}' \equiv C^{-1}\mathbf{e}$ and $\mathbf{f}' \equiv C^{-1}\mathbf{f}$. To get a feeling for what is going on, let us study a simple special case. If the points $\{\mathbf{k}_i\}$ are all separated by several coherence lengths, $C$ will be approximately diagonal. If we furthermore have chosen all the $\mathbf{k}_i$ to lie near a spherical shell $|\mathbf{k}| = k_*$ and the survey geometry is spherically symmetric, then we can choose units where $C = I$, the identity matrix, and the solution reduces to

$$w_i = \frac{1}{N} + \left(\frac{1}{N}\sum_{j=1}^{N} f_j\right) - f_i. \tag{38}$$

Hence the weights are just a constant minus the window-convolved penalties $f_i$. If some $f_i$ exceeds the average by more than $1/N$, $w_i$ will become negative, which is forbidden. If this happens, we clearly get a reasonably good result by simply setting all negative weights equal to zero and rescaling. We get an even better result if we reduce $N$ by omitting these offending $\mathbf{k}$-values and solving again, a procedure which is easy to iterate until all weights are positive. This procedure is of course merely a useful numerical trick which produces an almost optimal solution with little numerical effort — if the exact optimum is desired, then a standard software package for quadratic programming should be employed instead. It should be stressed that even with an approximate solution for the weights, the resulting power spectrum estimate is of course 100% correct — the resulting error bars will merely be slightly larger than the optimal ones.

# 5  The pixelization method

The problem of estimating $P(k)$ from galaxy surveys has many elements in common with that of estimating the angular power spectrum $C_\ell$ from cosmic microwave background (CMB) sky maps – see Tegmark (1994, hereafter T94), Tegmark (1995) and references therein. One obvious difference is that CMB maps are pixelized and galaxy surveys are not. In this section, we discretize the galaxy survey problem by dividing the sky into $N$ "fuzzy pixels". This allows the calculations of the two preceding sections to be combined into a single eigenvalue problem. The main advantage is that the eigenvalue problem is no longer one of differential operators but one of finite matrices. Thus it can be always be solved numerically even when we add in various extra complications, such as for instance redshift distortions, evolution and power-spectrum weighting, as outlined at the end of the section. The main disadvantage is that since $N \times N$ matrices become numerically cumbersome to diagonalize if $N$ exceeds a few thousand, we must make our 3D pixels fairly wide and thus sacrifice some spatial resolution. This makes the method most suitable for extracting the power at long wavelengths, where spatial resolution is not a problem.

Let us define the overdensity in $N$ "pixels" $z_1, ..., z_N$ by

$$z_i \equiv \int \varphi_i(\mathbf{r})[n(\mathbf{r}) - \bar{n}(\mathbf{r})]d^3r. \tag{39}$$

Although the weight functions $\varphi_i$ may be chosen to live only on disjoint volumes (sharp pixelization, in which case the pixel values reduce to being basically counts in cells), it is preferable to reduce sidelobes in Fourier space by chosing the $\varphi_i$ to be smooth, for instance Gaussians $\varphi_i(\mathbf{r}) \propto e^{-|\mathbf{r}-\mathbf{r}_i|^2/2\sigma_i^2}$. It is convenient to chose the widths $\sigma_i$ to be slightly greater than the typical separation between neighboring points $\mathbf{r}_i$, and to place points more sparsely in the outskirts of the survey volume.

We define the matrix-valued functions $S(\mathbf{k})$ and $T(\mathbf{r})$ by $S_{ij}(\mathbf{k}) \equiv \widehat{\psi}_i(\mathbf{k})^*\widehat{\psi}_j(\mathbf{k})$ and $T_{ij}(\mathbf{r}) \equiv \psi_i(\mathbf{r})^*\psi_j(\mathbf{r})$, where $\psi_i(\mathbf{r}) \equiv \bar{n}(\mathbf{r})\varphi_i(\mathbf{r})$. The rest of the analysis now becomes very similar to that in T94, so in the interest of brevity, we will simply copy that notation and refer to Section III of that paper for explanations. The power estimate

$$\tilde{D}(k^*) \equiv \mathbf{z}^t E \mathbf{z} = \sum_{k=1}^{N'} \alpha_k (\mathbf{e}_k \cdot \mathbf{z})^2 \tag{40}$$

has the expectation value

$$\langle \tilde{D}(k^*) \rangle = \int W(\mathbf{k})P(\mathbf{k})d^3k + B, \qquad (41)$$

where the window function is

$$W(\mathbf{k}) = \text{tr}\,\{ES(\mathbf{k})\} = \sum_{k=1}^{N'} \alpha_k \mathbf{e}_k^t S(\mathbf{k}')\mathbf{e}_k \qquad (42)$$

and the bias from shot noise is

$$B = \sum_{k=1}^{N'} \alpha_k \mathbf{e}_k^t \left[\int T(\mathbf{r})V(\mathbf{r})d^3r\right] \mathbf{e}_k. \qquad (43)$$

Minimizing $\int p(k)W(\mathbf{k})d^3k + B$ (note that the penalty function $p$ depends only on the magnitude $k = |\mathbf{k}|$) subject to the constraint that $\int W(\mathbf{k})d^3k = 1$, we obtain the generalized eigenvalue problem

$$(Q - \lambda R)\mathbf{e}_k = 0, \qquad (44)$$

where we have defined the $N \times N$ matrices

$$Q = \int S(\mathbf{k})p(k)d^3k + \int T(\mathbf{r})V(\mathbf{r})d^3r, \qquad (45)$$

$$R = \int S(\mathbf{k})d^3k. \qquad (46)$$

Since both $Q$ and $R$ are Hermitean ($T$ is in addition real), this generalized eigenvalue problem has $N$ orthogonal solutions, which we normalize and sort by their eigenvalues (the smallest eigenvalue corresponds to $k = 1$, the best solution, *etc.*). Most standard eigenvalue packages (such as the public-domain package EISPACK, or that included in NAG) provide a specialized routine for precisely this problem: the generalized eigenvalue problem where $Q$ and $R$ are Hermitean, and $R$ is positive definite. Now the trade-off between vertical and horizontal error bars becomes quite transparent. The larger we choose $N'$, *i.e.*, the more modes we include (with equal weighs $\alpha = 1/N'$, say), the smaller the variance of our power estimate $\tilde{D}(k^*)$ and the wider the window function $W(k) = \int W(\mathbf{k})d\Omega_k$. Since the modes are sorted by decreasing merit, we can for instance plot vertical and horizontal error bars versus $N'$, the number of modes included, and then fix $N'$ according to our preferences, analogously to what is done in a Karhunen-Loève analysis à la V94.

## 5.1 Redshift distortions

A ubiquitous problem with power spectrum estimation is that of "redshift distortions". When estimating the radial distance to a galaxy by its redshift, galaxies receding faster than the Hubble flow due to local gravitational interactions appear to be further away than they really are, and vice versa. This was first discussed by Kaiser (1987), and a recent review is given by Tegmark & Bromley (1995). Denoting the apparent density field $\delta_s(\mathbf{r})$, it is straightforward to use Kaiser's formalism to show that in linear perturbation theory,

$$\widehat{\delta}_s(\mathbf{k}) = \widehat{\delta}_r(\mathbf{k}) + \beta \int f(\mathbf{k}, \mathbf{k}') \widehat{\delta}_r(\mathbf{k}') d^3k', \qquad (47)$$

where $\beta \equiv \Omega^{0.6}/b$, the constant $b$ is the so called bias factor, and the function $f$ is defined by

$$f(\mathbf{k}, \mathbf{k}') \equiv e^{i(\mathbf{k}'-\mathbf{k})\cdot\mathbf{r}} \left[ (\widehat{\mathbf{k}}' \cdot \widehat{\mathbf{r}})^2 - \frac{2i}{k'r}(\widehat{\mathbf{k}}' \cdot \widehat{\mathbf{r}}) \right] d^3r. \qquad (48)$$

Thus we obtain

$$\langle \widehat{\delta}_s(\mathbf{k})^* \widehat{\delta}_s(\mathbf{k}') \rangle = \int g(\mathbf{k}, \mathbf{k}', \mathbf{k}'') P(\mathbf{k}'') d^3k'', \qquad (49)$$

where

$$\begin{aligned} g(\mathbf{k}, \mathbf{k}', \mathbf{k}'') &\equiv \delta(\mathbf{k} - \mathbf{k}'')\delta(\mathbf{k}' - \mathbf{k}'') + \beta^2 f(\mathbf{k}, \mathbf{k}'')^* f(\mathbf{k}', \mathbf{k}'') + \\ &+ \beta \left[ \delta(\mathbf{k} - \mathbf{k}'') f(\mathbf{k}', \mathbf{k}) + \delta(\mathbf{k}' - \mathbf{k}'') f(\mathbf{k}, \mathbf{k}')^* \right]. \end{aligned} \qquad (50)$$

The above expressions are derived and discussed in detail by Zaroubi & Hoffman (1995). The key point here is that although no longer diagonal, and rather messy, the expression for $\langle \widehat{\delta}_s(\mathbf{k})^* \widehat{\delta}_s(\mathbf{k}') \rangle$ is still linear in the power spectrum. Thus the expectation value of a quadratic estimator will still be some noise term plus a term *linear* in $P(\mathbf{k})$. In other words, if the treatment in the previous sections is repeated with $\delta_s$ in place of $\delta_r$, all the optimization problems will retain the same form, merely with messier looking expressions for the window functions — window functions that now probe the power directly in real space, not in redshift space. Although the resulting analog of equation (5) will generally become too complex to admit analytic solutions, the extra complication is of no importance in the case of equation (44) above, since the latter is to be solved numerically anyway. All that happens is that the matrices to be diagonalized change somewhat. Since all linear complications of the problem are numerically "free", we mention below two more elements of realism that are straightforward to take into account.

## 5.2 Evolution

If smoothed on a scale much greater than the Jeans length, the field of density fluctuations $\delta_r$ maintains its shape in linear perturbation theory, simply increasing in amplitude by a position-independent growth factor $D$. Since we are seeing distant galaxies at an earlier time, we see the apparent density fluctuations

$$\delta_a(\mathbf{r}) \equiv D(z)\delta_r(\mathbf{r}), \tag{51}$$

where the redshift $z = H_0 r/c$ if $z \ll 1$ and $D(z) = 1/(1+z)$ for the simple case $\Omega = 1$. Equation (3) now gets replaced by

$$\langle |\widehat{F}(\mathbf{k}^*)|^2 \rangle = |\langle \widehat{F}(\mathbf{k}^*) \rangle|^2 + \int |\widehat{\psi}(\mathbf{k})|^2 P(\mathbf{k}) d^3k + \int \frac{|\psi(\mathbf{r})|^2}{D(r)^2 \bar{n}(\mathbf{r})} d^3r, \tag{52}$$

where we have redefined $\psi(\mathbf{r}) = D(r)\bar{n}(\mathbf{r})\varphi(\mathbf{r})$. This correction is quite small for contemporary galaxy surveys, where $\bar{n}$ typically varies dramatically between $z=0$ and $z=0.2$, a range over which $D$ changes by at most about 20%, less for small $\Omega$.

This fluctuation evolution should not be confused with galaxy evolution, which affects only $\bar{n}$ and not $\delta_r$.

## 5.3 Power weighting

Our estimators probe a weighted average of $P(k)$. This means that in regions where $P(k)$ is concave, i.e., $P''(k) > 0$, we get an overestimate, and vice versa. In general, averaging with a window function causes the least harm when applied to a function that is fairly constant, since then "leakage" from the sides has little impact. If we have certain preconceptions about the behavior of $P$, believing it to have roughly the form $P_*(k)$, say, is thus better to define a rescaled power spectrum $D(k) \equiv P(k)/P_*(k)$ and set out to estimate the fairly constant function $D(k)$ instead. This is precisely what was done with the CBR spectrum in T94, and we will not repeat it here as it is completely analogous. The result is that the Schrödinger eigenvalue problem (5) gets replaced by a generalized eigenvalue problem ($E$ gets multiplied by an operator) which is harder to solve. The discrete eigenvalue problem in equation (44) is already of the generalized type, so this is no complication at all. For instance, when going for the power on very large scales, one may chose $P_*(k) = k$ to make the window functions shun high frequencies more than they would otherwise.

# 6 Discussion

We have presented a method for estimating the 3D power spectrum $P(k)$ that maximizes the spectral resolution, *i.e.*, makes the window functions in $k$-space as narrow as possible given a level of shot-noise variance. The method decomposes into two separate steps:

1. Estimate $P(\mathbf{k})$ for a 3D grid of $\mathbf{k}$-values with separations of the order of the coherence length.
2. For each $k$-value of interest, estimate $P(k)$ by taking a weighted average of the estimates of $P(\mathbf{k})$.

We found that Step 1 amounted to the the following:

- Determine the weight function $\psi_0(\mathbf{r})$ from the survey geometry by finding the ground state solution of the Schrödinger equation with the inverse selection function as potential.
- Weight the galaxies by the inverse selection function times $\psi_0$.
- Fourier transform. (This simply involves a sum over all the observed galaxies.)
- As the power estimate, take the square modulus of the Fourier transform minus $|\widehat{\psi}_0(\mathbf{k})|^2$ and the $\mathbf{k}$-independent noise bias $\int \frac{\psi_0^2}{\bar{n}} d^3r$.

It is noteworthy that whereas the second step does, this first step does *not* involve any assumptions about the power spectrum $P(k)$.

For Step 2, the standard approach of giving equal weight to all $\mathbf{k}$-values on a spherical shell of radius $k$, and zero weight to all others, gives a decent estimate if the survey volume is fairly spherical. In general, the optimal weighs are obtained by inverting a certain matrix as described in Section 4. For pencil-beams and slices, those $\mathbf{k}$-values near the shell that get the greatest weight are those that correspond to the widest directions of the survey volume, just as one would expect.

From the discussion at the end Section 3, we conclude that if a fixed number of square degrees are available for a redshift survey, then

- one should not try to make the area oblong, in a naive attempt to "probe larger scales",
- one should not split the area into several disjoint pieces, and
- the optimal survey shape is a circle, corresponding to a conical 3D geometry.

In Section 5, we presented a method where the survey volume was divided into a large number of "fuzzy pixels". This discretization of the problem makes it quite analogous to the analysis of pixelized cosmic microwave background (CMB) sky maps, and the solution is of course the same as well: the optimal power estimators are eigenvectors of a matrix eigenvalue problem, just as for the CMB case treated in T94. Because of the limited spatial resolution of the pixels, this method is most suited for estimating the power at the longest wavelengths. A major advantage is that it offers a way of taking linear regime redshift distortions into account without any approximations.

Although it is hoped that the methods presented in this paper will prove useful in estimating the amount of power on the largest scales, it should be stressed that wide window functions are by no means the only challenge we face in this endeavor. One notorious difficulty is that $\bar{n}$ is usually not known a priori, but estimated from the survey itself, which can gives the impression of artificially low large-scale power as discussed in FKP. Another potential source of trouble, for flux-limited surveys, is that the bias parameter may depend on the luminosity class of galaxies under consideration. Since the most remote parts of the survey probe mainly the brightest galaxies, the observed $\delta_r$ would have a spatial modulation. Although the sign of this modulation is probably the opposite of the evolution effect discussed in the previous section, the net result would be the same: an overestimate of the large-scale power.

Let us conclude by comparing the results from Section 2 with those of FKP. They derive the weight function that minimizes the vertical error bars, the variance of the estimate, without specific regard to what happens to the window function. With our notation, they find the optimal solution to be

$$\psi_0(\mathbf{r}) \propto \frac{\bar{n}(\mathbf{r})}{1 + \bar{n}(\mathbf{r})P(k)}. \tag{53}$$

If $\bar{n}(\mathbf{r})P(k) \gtrsim 1$ in the central regions of the survey, this implies that their optimal $\psi_0$ is fairly constant there, which agrees with our conclusions: the ground state of the Schrödinger equation does not vary much in regions far from the survey boundaries. Thus the powerful conclusions that can be drawn from the FKP formula about the merits of sparse sampling, for instance, remain valid also with the analysis method presented here. Rather, the difference between the two methods manifests itself near the edges, where our $\psi_0$ always approaches zero faster than the $\psi_0$ of FKP. The latter falls off no faster than $\bar n$ does (for instance, it goes abruptly to zero at any sharp boundary). This is why it gives a wider window function in **k**-space.

The author wishes to thank George Efstathiou, Karl Fisher, Alan Heavens, John Peacock and Michael Vogeley for useful comments on the manuscript, and Wayne Hu and Naoshi Sugiyama for kindly providing unpublished power spectra. This work has been partially supported by European Union contract CHRX-CT93-0120 and Deutsche Forschungsgemeinschaft grant SFB-375.

# Appendix

In this Appendix, we derive equations (2), (3), (28) and (29).

As mentioned, we model the galaxy distribution as a 3D stochastic point process $n(\mathbf{r}) = \sum_i \delta(\mathbf{r} - \mathbf{r}_i)$ which is a Poisson process with intensity (average point density) $\lambda(\mathbf{r})$. A Poisson process satisfies (see *e.g.* Appendix A of FKP)

$$\langle n(\mathbf{r}) \rangle = \lambda(\mathbf{r}), \qquad (54)$$
$$\langle n(\mathbf{r})n(\mathbf{r}') \rangle = \lambda(\mathbf{r})\lambda(\mathbf{r}') + \delta(\mathbf{r} - \mathbf{r}')\lambda(\mathbf{r}). \qquad (55)$$

Here $\lambda$ is it self a random field, $\lambda(\mathbf{r}) = \bar n(\mathbf{r})[1+\delta_r(\mathbf{r})]$, where the density fluctuations $\delta_r$ are a Gaussian random field satisfying the standard expressions $\langle \delta_r(\mathbf{r}) \rangle = 0$ and[8]

$$\langle \widehat\delta_r(\mathbf{k})^* \widehat\delta_r(\mathbf{k}') \rangle = (2\pi)^6 \delta(\mathbf{k} - \mathbf{k}') P(\mathbf{k}). \qquad (56)$$

There are thus two separate random steps involved in generating $n$: first the generation of $\delta_r$, then the Poissonian distribution of points. To make this distinction clear, will will occasionally use double brackets, where the inner bracket denotes expectation values over $\delta_r$. For instance, $\langle\langle n(\mathbf{r}) \rangle\rangle = \langle \lambda(\mathbf{r}) \rangle = \bar n(\mathbf{r})$. Given a function $\varphi_0$, we define $F(\mathbf{r}) \equiv \varphi_0(\mathbf{r})n(\mathbf{r})$ and wish to compute the mean and autocorrelation of its Fourier transform

$$\widehat F(\mathbf{k}) \equiv \int e^{-i\mathbf{k}\cdot\mathbf{r}} F(\mathbf{r}) d^3r. \qquad (57)$$

Inserting the definition of $F$, we obtain

$$\langle\langle \widehat F(\mathbf{k}) \rangle\rangle = \int e^{-i\mathbf{k}\cdot\mathbf{r}} \varphi_0(\mathbf{r}) \langle\langle n(\mathbf{r}) \rangle\rangle d^3r = \int e^{-i\mathbf{k}\cdot\mathbf{r}} \psi_0(\mathbf{r}) = \widehat\psi_0(\mathbf{k}), \qquad (58)$$

---

[8] We adopt this slightly unconventional $2\pi$-convention in the definition of $P(\mathbf{k})$ because it eliminates the nuisance of repeated occurrences of $(2\pi)^3$ throughout the main part of the paper.

where $\psi_0(\mathbf{r}) \equiv \bar{n}(\mathbf{r})\varphi_0(\mathbf{r})$ as before, and

$$\langle\langle \widehat{F}(\mathbf{k})^* \widehat{F}(\mathbf{k}') \rangle\rangle = \int\int e^{-i(\mathbf{k}'\cdot\mathbf{r}'-\mathbf{k}\cdot\mathbf{r})} \varphi_0(\mathbf{r})^*\varphi_0(\mathbf{r}') \langle\langle n(\mathbf{r})n(\mathbf{r}')\rangle\rangle d^3r\, d^3r'. \quad (59)$$

The contribution to $\langle\langle n(\mathbf{r})n(\mathbf{r}')\rangle\rangle$ from the second term in equation (55) gives

$$\int e^{-i(\mathbf{k}'-\mathbf{k})\cdot\mathbf{r}} |\varphi_0(\mathbf{r})|^2 \bar{n}(\mathbf{r}) d^3r = \int e^{-i(\mathbf{k}'-\mathbf{k})\cdot\mathbf{r}} \frac{|\psi_0(\mathbf{r})|^2}{\bar{n}(\mathbf{r})} d^3r, \quad (60)$$

which we identify as the shot noise contribution $C_s$ in equation (29). The contribution to $\langle\langle n(\mathbf{r})n(\mathbf{r}')\rangle\rangle$ from the first term in equation (55) gives

$$\int\int e^{-i(\mathbf{k}'\cdot\mathbf{r}'-\mathbf{k}\cdot\mathbf{r})} \psi_0(\mathbf{r})\psi_0(\mathbf{r}') \left[1 + \langle\delta_r(\mathbf{r})\delta_r(\mathbf{r}')\rangle\right] d^3r\, d^3r'. \quad (61)$$

The first of these terms is separable and reduces to simply $\langle\langle \widehat{F}(\mathbf{k})\rangle\rangle^* \langle\langle \widehat{F}(\mathbf{k})\rangle\rangle$. Both $\psi_0$ and $\delta_r$ are real-valued, so using the Fourier identity

$$\int e^{-i\mathbf{k}\cdot\mathbf{r}} \psi_0(\mathbf{r})^* \delta_r(\mathbf{r}) d^3r = \frac{1}{(2\pi)^3} \int \widehat{\psi}_0(\mathbf{k}''-\mathbf{k})^* \widehat{\delta}_r(\mathbf{k}'') d^3k'', \quad (62)$$

the second term yields

$$\frac{1}{(2\pi)^6} \int\int \widehat{\psi}_0(\mathbf{k}'''-\mathbf{k}')^* \widehat{\psi}_0(\mathbf{k}''-\mathbf{k}) \langle\widehat{\delta}_r(\mathbf{k}'')^* \widehat{\delta}_r(\mathbf{k}''')\rangle d^3k''\, d^3k'''$$

$$= \int \widehat{\psi}_0(\mathbf{k}''-\mathbf{k}')^* \widehat{\psi}_0(\mathbf{k}''-\mathbf{k}) P(\mathbf{k}'') d^3k'', \quad (63)$$

which is the cosmic contribution $C_c$ in equation (28). This completes the derivation of the covariance function $C(\mathbf{k},\mathbf{k}')$. The special case of equation (3) is obtained by setting $\mathbf{k} = \mathbf{k}' = \mathbf{0}$.

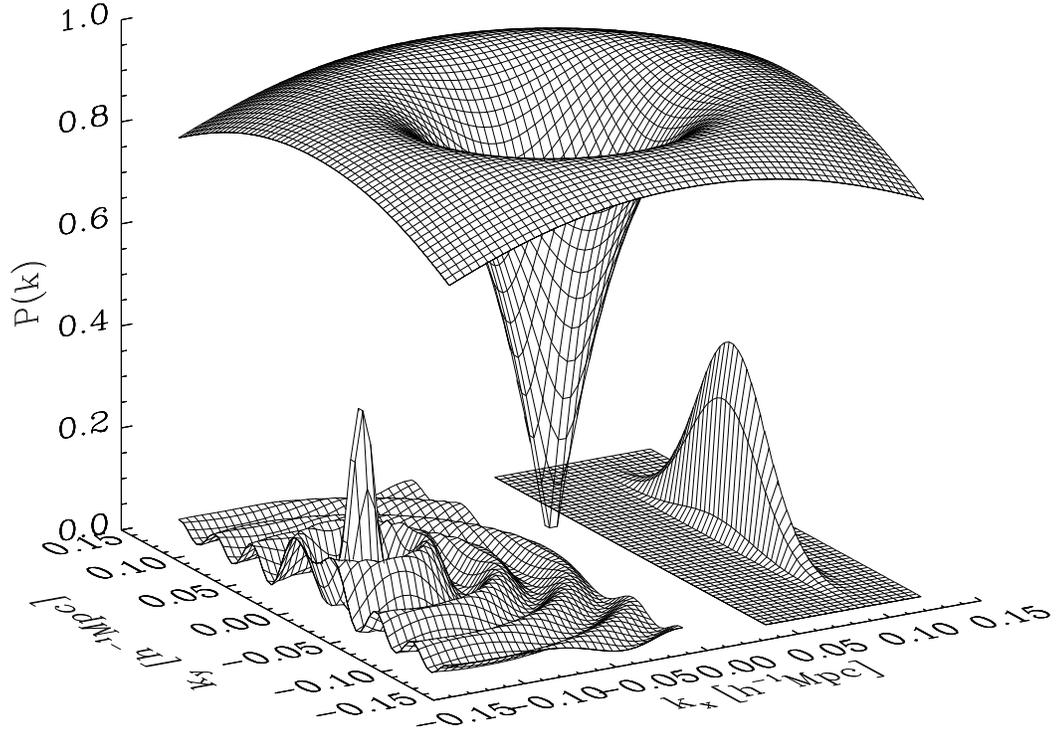

Figure 1: Problems in **k**-space.
The funnel-shaped surface is a standard CDM power spectrum $P(k_x, k_y, k_z)$, with $k_z = 0$. The window-function on the lower left suffers from wide sidelobes due to a sharp edge in the weight function $\psi_0$. The window function on the lower right suffers from being highly anisotropic, such as is the case for pencil beam and slice surveys.

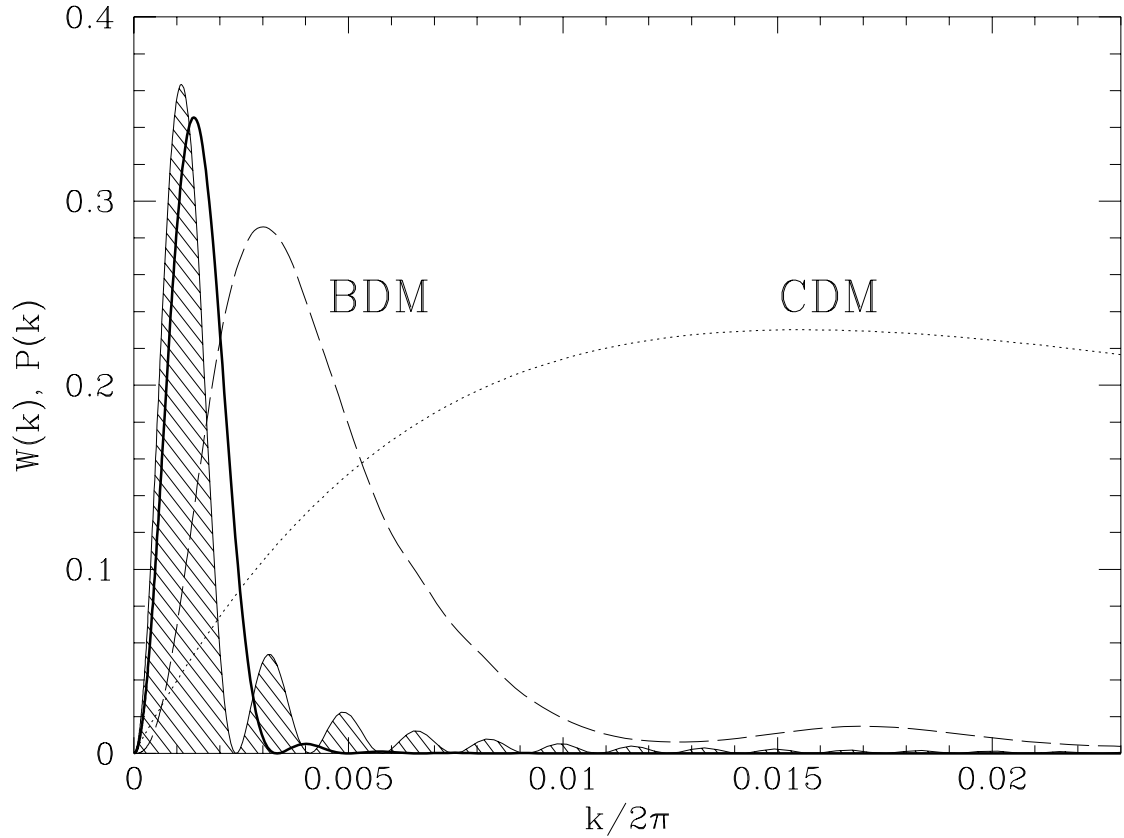

Figure 2: Window functions before and after optimization. The two solid lines show window functions for estimating the power at $k = 0$ in a volume limited survey of depth $300h^{-1}$Mpc. The heavy line corresponds to the optimal choice of $\psi_0$, whereas the one with all the sidelobes (shaded) corresponds to naively chosing $\psi_0$ constant. Two popular power spectra, BDM (dashed) and CDM (dotted) are plotted for comparison.